\documentclass[a4paper]{article}
\usepackage[top=2.8cm,bottom=3cm,left=3cm,right=3cm]{geometry}
\usepackage{graphicx}
\usepackage{txfonts}
\usepackage{amsfonts}
\usepackage{amssymb}
\usepackage{booktabs}
\usepackage{xcolor}
\usepackage{lineno}
\usepackage{natbib}
\bibpunct{(}{)}{;}{a}{}{,} % to follow the A&A style

\begin{document}
\Large
{\textbf{{\begin{center}Deep Learning for Active Region Classification: \\A Systematic Study from Convolutional Neural Networks to Vision Transformers\end{center}}}

\normalsize

Edoardo Legnaro$^1$, Sabrina Guastavino$^{1,2}$, Michele Piana$^{1,2}$, and Anna Maria Massone$^1$ \\

\hspace{-0.5cm}$^1$ MIDA, Dipartimento di Matematica, Universit\`a di Genova, via Dodecaneso 35 16146 Genova, Italy  \\
$^2$ Istituto Nazionale di Astrofisica, Osservatorio Astrofisico di Torino, via Osservatorio 20 10025 Pino Torinese Italy \\

\date{today}

%% Note that the \and command from previous versions of AASTeX is now
%% depreciated in this version as it is no longer necessary. AASTeX 
%% automatically takes care of all commas and "and"s between authors names.

%% AASTeX 6.3 has the new \collaboration and \nocollaboration commands to
%% provide the collaboration status of a group of authors. These commands 
%% can be used either before or after the list of corresponding authors. The
%% argument for \collaboration is the collaboration identifier. Authors are
%% encouraged to surround collaboration identifiers with ()s. The 
%% \nocollaboration command takes no argument and exists to indicate that
%% the nearby authors are not part of surrounding collaborations.

%% Mark off the abstract in the ``abstract'' environment. 
\begin{abstract}
    A solar active region can significantly disrupt the Sun–Earth space environment, often leading to severe space weather events such as solar flares and coronal mass ejections. As a consequence, the automatic classification of active region groups is the crucial starting point for accurately and promptly predicting solar activity. This study presents our results concerned with the application of deep learning techniques to the classification of active region cutouts based on the Mount Wilson classification scheme. Specifically, we have explored the latest advancements in image classification architectures, from Convolutional Neural Networks to Vision Transformers, and reported on their performances for the active region classification task, showing that the crucial point for their effectiveness consists in a robust training process based on the latest advances in the field.
    %determine the most effective network architectures for the sunspot classification task.
\end{abstract}

\textbf{key words.} Active region classification; deep learning; transformers

\section{Introduction}

The Sun is the source of a variety of dynamic phenomena, such as solar flares and coronal mass ejections (CMEs), which significantly impact space weather. These solar activities can cause disturbances that affect satellite operations, power grids, and communication systems, potentially leading to significant economic and technological disruptions \citep{baker2002cope, buzulukova2022space}. Accurate forecasting of space weather events is therefore crucial to mitigate their adverse effects on modern infrastructure \citep{guastavino2022implementation, guastavino2023operational, georgoulis2024prediction}. 

It is well-established that solar flares, i.e. intense bursts of radiation originating in the solar atmosphere, predominantly above sunspots \citep{hudson2011global, benz2017flare,piana2022hard}, are the main trigger of space weather. Therefore, the prediction of solar flares has become one of the most fundamental tasks of space weather understanding, and the development of data-driven models relying on artificial intelligence (AI) approaches is disruptively changing the scenario of this scientific and technological challenge \citep{guastavino2022implementation, guastavino2023operational}. Recent studies have shown a strong correlation between the likelihood of solar flare occurrences and specific characteristics of active region groups, including their size, the number of sunspots, and their magnetic classification \citep{mccloskey2018flare,eren2016flare}. It follows that active region classification represents the crucial initial step in predicting space weather.

Traditionally, expert observers manually classify active regions by examining images of the Sun, identifying sunspots, and categorizing them based on their size, shape and magnetic properties. This method has the disadvantages of being time-consuming, subjective, and not scalable with data volume. Nevertheless, the rapid accumulation of solar observations from an increasing number of space missions makes it possible to automating the classification of active region groups using machine learning techniques \citep{colak2008automated, abd2010automated, pesnell2012solar, scherrer2012helioseismic}, and recent advancements in the field, particularly deep learning \citep{lecun2015deep, goodfellow2016deep}, have shown promising results in this task \citep{fang2019deep,tang2021multiple}. This study builds upon the research presented in \citep{fang2019deep} and \citep{tang2021multiple} to explore and critically compare various state-of-the-art deep learning architectures for image classification, specifically focusing on convolutional neural networks (CNNs) and Vision Transformers (ViTs) when applied to solar active region cutouts. Specifically, we evaluated models trained on magnetograms, continuum images, or both, using 2D convolutions, and progressively trained models of increasing complexity to determine which architecture is most robust for this classification task. This evaluation incorporates extensive on-the-fly data augmentation and presents results based on five-fold cross-validation.

The plan of the paper is as follows. Section 2 describes the properties of the data set used for the experiments, and the corresponding pre-processing step. Section 3 provides details on the design and properties of the deep learning networks employed for the analysis. Section 4 describes the results of the study. Our conclusions are offered in Section 5.

\section{Data}
\subsection{The Dataset}
This study considered the SOLAR-STORM1 dataset provided by the Space Environment Warning and AI Technology Interdisciplinary Innovation Working Group \citep{fang2019deep}\footnote{The dataset is available at https://tianchi.aliyun.com/dataset/dataDetail?dataId=74779}.
This dataset contains continuous and magnetogram images provided by the Spaceweather HMI Active Region Patch (SHARP) from the Helioseismic and Magnetic
Imager (HMI) on-board the Solar Dynamics Observatory \citep{pesnell2012solar,scherrer2012helioseismic}. The images in this dataset are sunspots cutouts of different aspect ratios, as shown in Figure \ref{fig: aspectratios}.

\begin{figure}
    \centering
    \includegraphics[width = 0.6\textwidth]{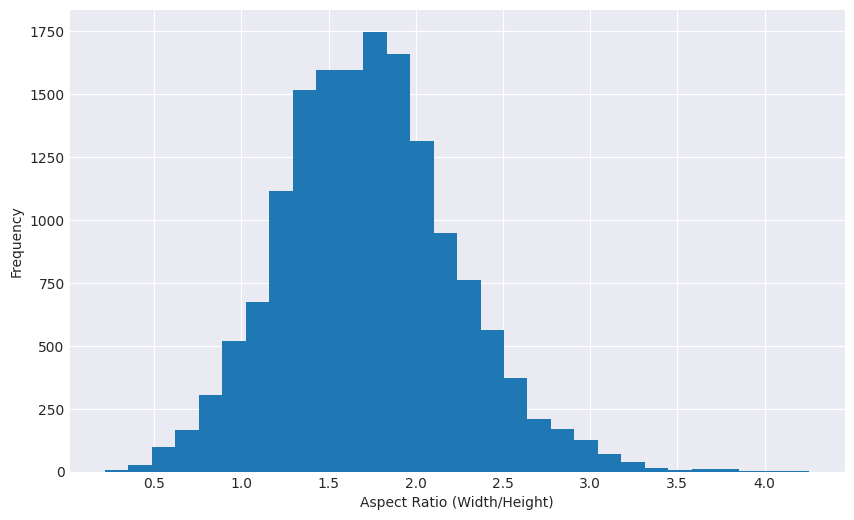}
    \caption{Histogram displaying the distribution of aspect ratios (width/height) of the images in the SOLAR-STORM1 dataset. }
    \label{fig: aspectratios}
\end{figure}

The temporal range of the dataset covers the interval from 2010 May 1 to 2017 December 12 (see Figure \ref{fig:kfold_split_visualization}, top panel) and the classes distribution is shown in Table \ref{tab:class_distribution}, second column. 
Specifically, the Mount Wilson sunspot magnetic types are divided into the following 8 classes.

\begin{itemize}
    \setlength\itemsep{0.01em}
    \item Alpha: Unipolar sunspot group.
    \item Beta: Bipolar sunspot group with distinct positive and negative magnetic polarities.
    \item Gamma: Complex region with irregular positive and negative polarities, preventing bipolar classification.
    \item Delta: Umbrae with opposite polarities within one penumbra, separated by less than 2°.
    \item Beta–Gamma: Bipolar group with complex polarity distribution, lacking a clear dividing line.
    \item Beta–Delta: Beta group containing one or more Delta spots.
    \item Beta–Gamma–Delta: Beta–Gamma group containing one or more Delta spots.
    \item Gamma–Delta: Gamma group containing one or more Delta spots.
\end{itemize}

Due to the limited number of examples in some groups, the Delta and Gamma-Delta magnetic types were excluded, and the Beta–Gamma, Beta–Delta, and Beta–Gamma–Delta classes were merged into a single class, referred to as Beta-X. As a result, the problem was approached as a multi-class classification task with three categories: Alpha, Beta, and Beta-X. In this dataset, images were taken every 96 minutes, but a sunspot group does not change consistently within 24 hours (see \cite{mccloskey2016flaring}). Therefore, this dataset presents a high degree of redundancy, and a random split of the training, validation, and test sets (as done in \cite{fang2019deep}) would result in inflated test scores. 
As the dataset lacks active region identifiers, such as NOAA numbers, the only viable option was to apply a temporal split to ensure that highly similar images are not distributed across different sets.
This approach mirrors the splitting strategy used in \cite{tang2021multiple}, where all data after 2016 January 1 were designated as the test set, and the remaining data were used for fivefold cross-validation.
A visualization of this splitting method is presented at the bottom of Figure \ref{fig:kfold_split_visualization}. The number of classes along with their respective percentages for the training and validation sets across the five folds are detailed in Table \ref{tab:fold_distribution}, while the distribution for the test set is shown in Table \ref{tab:class_distribution}, third column.

\begin{table}
    \centering
    \begin{tabular}{ccc}
        \toprule
        \textbf{Class} & \textbf{Full Dataset Counts (\%)} & \textbf{Test Dataset Counts (\%)} \\
        \midrule
        Alpha   & 5276 (33.73\%)  & 567 (48.38\%)  \\
        Beta    & 7849 (50.18\%)  & 496 (42.32\%)  \\
        Beta-X  & 2516 (16.09\%)  & 109 (9.30\%)   \\
        \midrule
        \textbf{Total} & \textbf{15641} & \textbf{1172} \\
        \bottomrule
    \end{tabular}
    \caption{Distribution of class cardinality and rate in the whole SOLAR-STORM1 dataset and in the test set.}
    \label{tab:class_distribution}
\end{table}

\begin{table}
    \centering
    \small{
    \begin{tabular}{ccccccccc}
        \cmidrule(lr){2-5} \cmidrule(lr){6-9}
        & \multicolumn{4}{c}{\textbf{Train}} & \multicolumn{4}{c}{\textbf{Val}} \\
        \cmidrule(lr){2-5} \cmidrule(lr){6-9}
         & \textbf{Alpha (\%)} & \textbf{Beta (\%)} & \textbf{Beta-X (\%)} & \textbf{Total} & \textbf{Alpha (\%)} & \textbf{Beta (\%)} & \textbf{Beta-X (\%)} & \textbf{Total} \\
        \toprule
        Fold 1    & 3821 (33.01\%) & 5691 (49.17\%) & 2063 (17.82\%) & 11575 & 888 (30.68\%)  & 1662 (57.43\%) & 344 (11.89\%) & 2894 \\
        Fold 2    & 3845 (33.22\%) & 5547 (47.92\%) & 2183 (18.86\%) & 11575 & 864 (29.85\%)  & 1806 (62.40\%) & 224 (7.74\%)  & 2894 \\
        Fold 3    & 3992 (34.49\%) & 5849 (50.53\%) & 1734 (14.98\%) & 11575 & 717 (24.78\%)  & 1504 (51.97\%) & 673 (23.26\%) & 2894 \\
        Fold 4    & 3516 (30.38\%) & 6067 (52.41\%) & 1992 (17.21\%) & 11575 & 1193 (41.22\%) & 1286 (44.44\%) & 415 (14.34\%) & 2894 \\
        Fold 5    & 3662 (31.63\%) & 6258 (54.06\%) & 1656 (14.31\%) & 11576 & 1047 (36.19\%) & 1095 (37.85\%) & 751 (25.96\%) & 2893 \\
        \bottomrule
    \end{tabular}
    }
    \caption{Distribution of class cardinality and rate in the training and validation sets across folds.}
    \label{tab:fold_distribution}
\end{table}

\begin{figure}
    \centering
    \includegraphics[width=0.9\textwidth]{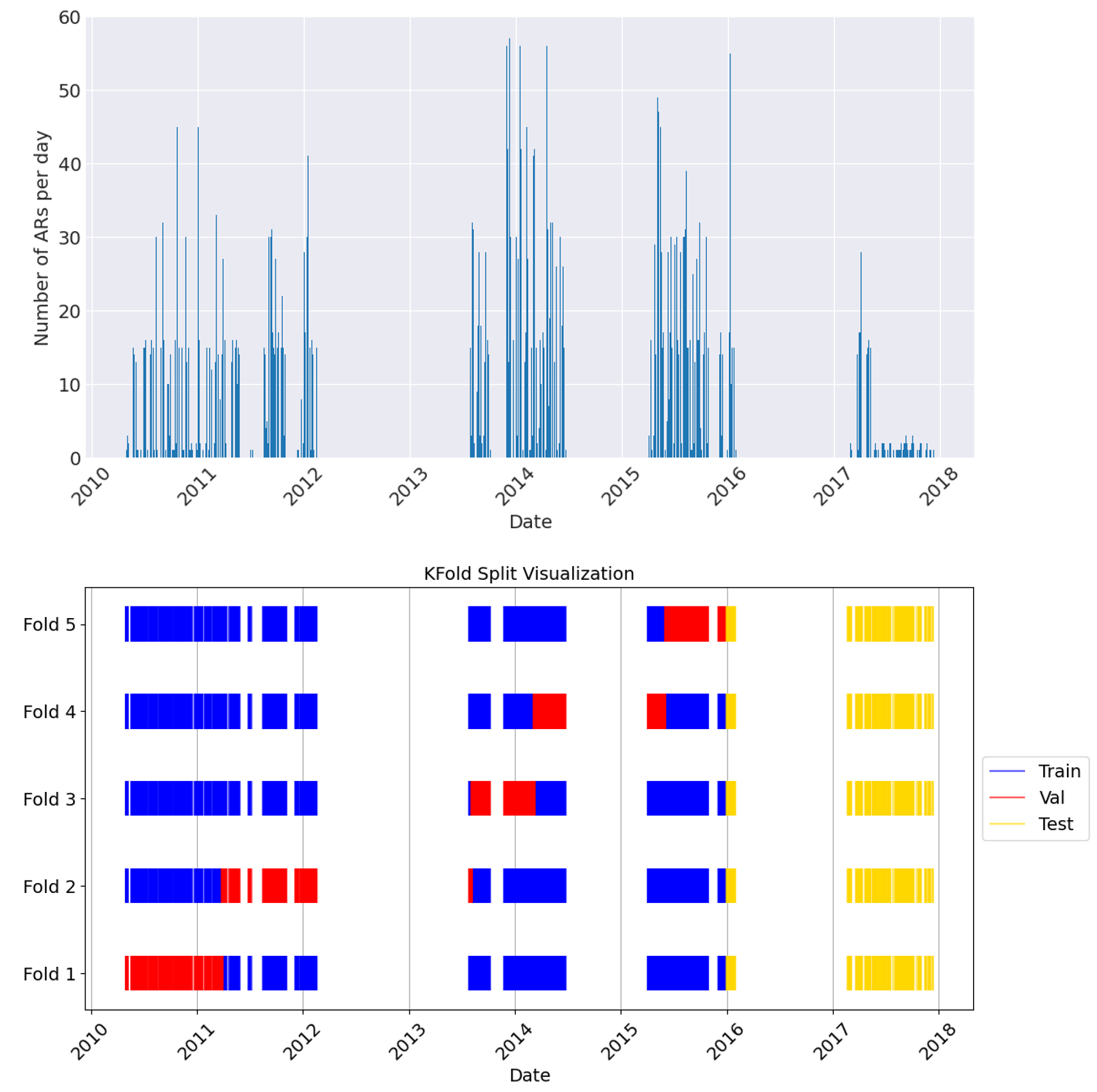}
    \caption{Temporal distribution of Active Regions (ARs) and the K-Fold split visualization for the dataset. 
    The upper panel shows the number of ARs per day present in the dataset, while the lower one illustrates how the data is divided into training (blue), validation (red), and test (yellow) sets across the five different folds.}
    \label{fig:kfold_split_visualization}
\end{figure}

\subsection{Data pre-processing}
Data augmentation is the process of artificially generating new data from existing one, primarily to train deep learning models, which require a large amount of training data samples in order to avoid overfitting. In this work the images are first normalized and cropped following the approach described in \cite{tang2021multiple}. In particular, magnetograms were normalized so that each pixel $m_{out}$ of the normalized image follows from
\begin{equation}
    \label{eq: HardTanh}
m_{out} = HardTanh\left(\frac{m_{in}}{d}\right) \  ,
\end{equation}
where  $m_{in}$ represents the original pixel value and $d\geq0$. The HardTanh function is defined as:
\begin{equation}
    HardTanh(x) = \left\{ \begin{array}{cc}
        1 &   x \ge 1 \\
        x &  -1 < x < 1\\
        -1 &  x \le -1\ \  , 
    \end{array}
    \right.
\end{equation}
which ensures that the final value lies within the interval $[-1, 1]$. In (\ref{eq: HardTanh}) we set  $d=800$, as done in \cite{fang2019deep} and \cite{tang2021multiple}, since this threshold value appears to give slightly higher performance metrics compared to higher constant values (see Table \ref{tab: divisors}). For continuum images, normalization is performed such that each pixel  $c_{out}$  is determined by
\begin{equation}
    \label{eq: MinMax}
    c_{out} = 1 - \frac{c_{in} - \min}{\max - \min},
\end{equation}
where $\min$ and $\max$ respectively denote the minimum and maximum values of the input image,  which has pixels $c_{in}$.
The final normalized values lie in the interval $[0,1]$, with background closer to 0 and the sunspots closer to 1.
Additionally, we included cropped images of the active regions in the dataset, using a strategy similar to that of \cite{tang2021multiple}. Specifically, this cropping was performed by comparing the absolute differences between the maximum values in the rows and columns of the normalized continuum images. Peaks in these differences were then identified, and a bounding box was selected by adding a margin around these peaks.
This is done in order to filter out noise, allowing the model to concentrate on the significant features relevant to the classification task.
An example of the normalization and cropping is shown in Figure \ref{fig: normalization}.
Next, we apply a random horizontal and/or vertical flip of the images, a perspective change, and an affine transformation including a minor translation and a rotation of a random angle up to $30^\circ$.
Usually, data augmentation techniques are applied before fitting a deep learning model on the whole dataset. 
However, in this way a single augmented dataset is created. Performing data augmentation on-the-fly as in \cite{cerqueira2024fly}, where a new set of random transformations is applied to the dataset at each epoch, creates slight variations with every iteration. Therefore, this approach enables better exploration of the data space and can help reduce the risk of overfitting.

\begin{figure}
    \centering
    \includegraphics[width=0.9\textwidth]{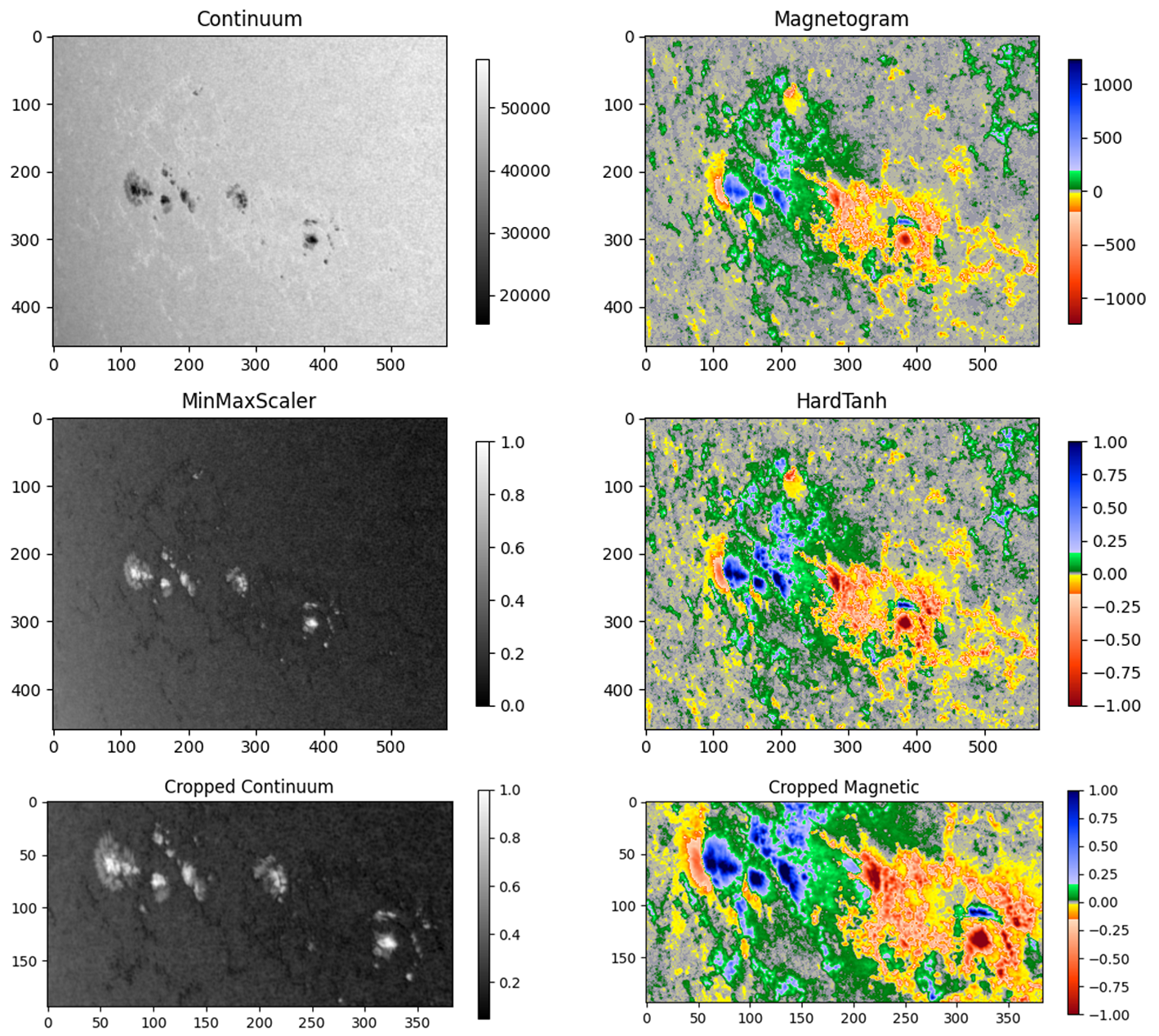}
    \caption{Effects of the normalization process on the images. The AR considered as an example is a Beta-X one at time 2014-03-17 16:00. The first row shows the original continuum and magnetogram images, while
    the second one shows, on the left, the normalized continuum image with Min-Max normalization (\ref{eq: MinMax}) and, on the right, the magnetogram image normalized using the HardTanh function (\ref{eq: HardTanh}) with constant $d=800$. The last line shows the cropping of the AR done to filter out noise and allow the model  to concentrate on the significant features relevant to the classification task.}
    \label{fig: normalization}
\end{figure}

\begin{figure}
    \centering
    \includegraphics[width=0.9\textwidth]{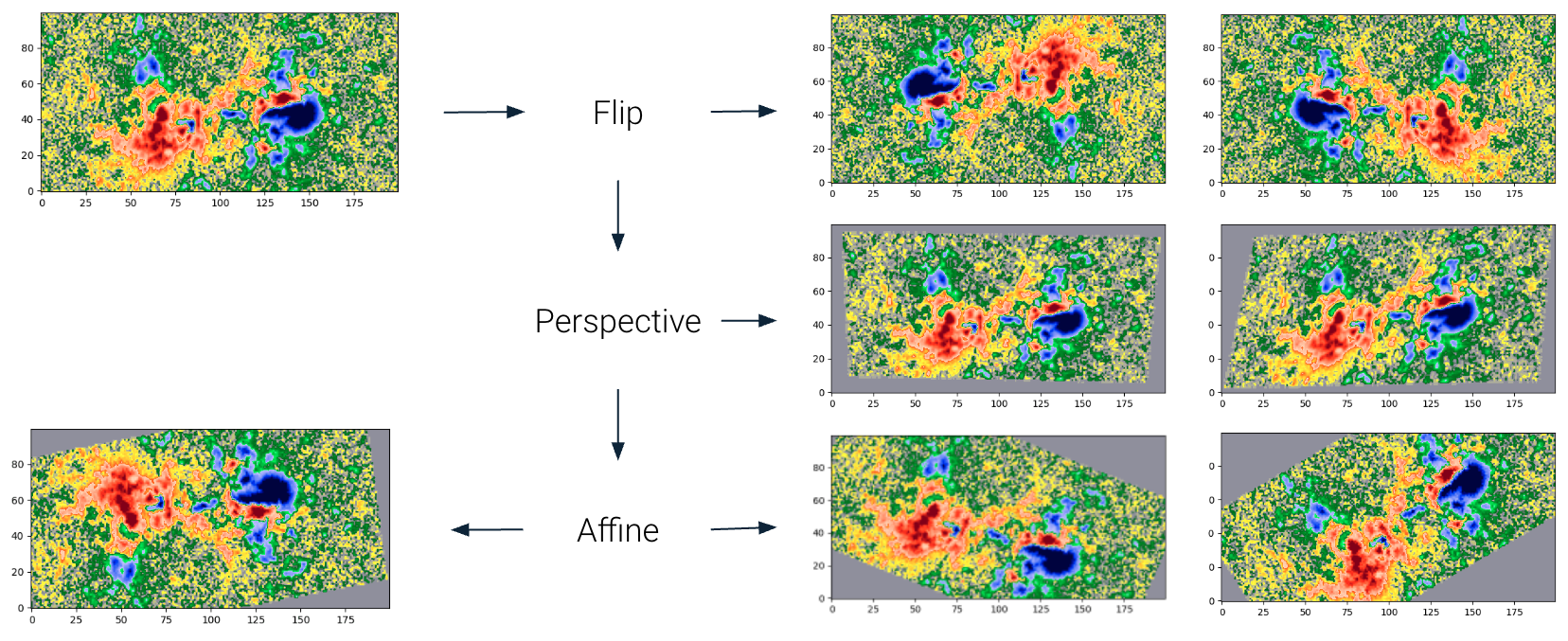}
    \caption{Illustration of data augmentation techniques. The original image (top left) undergoes a series of transformations to enhance the model’s ability to generalize. The first row shows the effect of horizontal and vertical flips. The second row demonstrates perspective transformations, altering the viewing angle. The third row applies affine transformations, including scaling, rotation, and shear, to further diversify the dataset.}
    \label{fig: augmentations}
\end{figure}

\begin{table}
    \centering
    \scriptsize{
    \begin{tabular}{cccccc}
    \toprule
    $d$ & Acc. ($\min$ - $\max$) & F1 Score ($\min$ - $\max$) & F1 Alpha ($\min$ - $\max$) & F1 Beta ($\min$ - $\max$) & F1 Beta-X ($\min$ - $\max$)\\ 
    \midrule
    800 & 0.8563 (0.8396 - 0.8805) & 0.8223 (0.7972 - 0.8472) & 0.9140 (0.8994 - 0.9295) & 0.8215 (0.7956 - 0.8551) & 0.7315 (0.6829 - 0.7570) \\
    1200 & 0.8466 (0.8191 - 0.8814) & 0.8048 (0.7659 - 0.8441) & 0.9159 (0.9103 - 0.9335) & 0.8089 (0.7750 - 0.8595) & 0.6895 (0.6115 - 0.7451) \\
    2000 & 0.8398 (0.8276 - 0.8532) & 0.7961 (0.7800 - 0.8079) & 0.9117 (0.8868 - 0.9241) & 0.7963 (0.7712 - 0.8161) & 0.6804 (0.6241 - 0.7424) \\
    4000 & 0.8493 (0.8106 - 0.8763) & 0.8104 (0.7657 - 0.8477) & 0.9135 (0.8881 - 0.9275) & 0.8111 (0.7641 - 0.8496) & 0.7066 (0.6449 - 0.7717) \\
    \bottomrule
    \end{tabular}
    }
    \caption{Performance metrics (Accuracy, Macro F1 Score (macro-averaged) and F1 Scores for the singles classe Alpha, Beta and Beta-X) for different divisor values ($d$). The values are presented as the average, minimum and maximum value across the five folds for each metric obtained on the test set.
    The model trained here is resnet18 on magnetograms, using the Adam optimizer with learning rate $10^{-5}$ and images resized to 224x224 squares.}
    \label{tab: divisors}
\end{table}

\section{Deep neural networks}
In this work, we explored different neural network architectures for classifying sunspots by their magnetic classes. To achieve this, we utilized the models provided in \cite{rw2019timm}.
The neural network architectures selected for this comparative study encompass both Convolutional Neural Networks (CNNs) and Vision Transformers (ViTs), since traditionally CNNs are better for capturing local features, while transformers focus more on global features. By comparing these two distinct architectures, this study aimed to evaluate their respective strengths and limitations in classifying solar active region cutouts, with the goal of determining the most robust approach for this classification task.

\subsection{Convolutional Neural Networks}
A Convolutional Neural Network (CNN) is a specific type of deep learning architecture designed for efficiently processing grid-like data, such as images. CNNs use convolutional layers to automatically extract hierarchical features by applying learned filters across the input data. Pooling layers further reduce spatial dimensions, helping to downsample feature maps and minimize computational complexity. Non-linear activation functions then introduce complexity into the model, allowing it to capture intricate data patterns. Finally, fully connected layers transform the learned features into a final prediction. CNNs excel at visual tasks due to their ability to learn spatial relationships and patterns directly from data \citep{lecun1998gradient, krizhevsky2012imagenet}. We now briefly introduce the CNNs whose performances have been explored in this paper.

\subsubsection{VGG}
The VGG (Visual Geometry Group) architecture, introduced in \cite{simonyan2014very}, consist in of sequential layers of small 3x3 convolutions stacked together, followed by max-pooling layers.
This design allows the network to capture increasingly complex features while maintaining manageable computational complexity. 

\subsubsection{Inception}
The Inception network, introduced by \cite{szegedy2015going} and also known as GoogLeNet, is a deep convolutional neural network involving the “Inception module,” which applies multiple convolutional filters of varying sizes (1x1, 3x3, and 5x5), and a pooling layer in parallel. This allows the network to capture features at different scales without significantly increasing the computational cost. Additionally, the architecture incorporates dimensionality reduction through 1x1 convolutions, which helps controlling the number of parameters and memory usage. 

\subsubsection{ResNet}
The ResNet (Residual Network) architecture, introduced in  \cite{he2016deep}, is a type of deep neural network that introduces connections or residual connections to address the vanishing gradient problem occurring in very deep neural networks. These connections bypass one or more layers by allowing the input of a layer to be added directly to its output, in such a way that the network passes information directly to deeper layers. This technique enables the model to learn identity mappings, making it easier to train deeper networks without degradation in performance.

\subsection{Vision Transformers}
Visual Transformers (ViTs) are a class of deep learning architectures designed to handle visual data using the transformer model, which was originally developed for natural language processing tasks \citep{dosovitskiy2020image}. Unlike CNNs, ViTs operate on image patches that are treated as a sequence of tokens, similar to words in a sentence, and are processed using self-attention mechanisms (see \cite{vaswani2017attention}). This mechanism is used to determine the importance of each element in a sequence relative to others. It allows the model to focus on relevant parts of the input data when generating a representation for each element.
In self-attention, for each input token (or element), the model computes three vectors: query, key, and value. The query from one token is compared with the keys of all other tokens to compute attention scores, which indicate how much focus the model should place on each token. These scores are then used to compute a weighted sum of the value vectors, creating a new representation for each token based on its relationships to others. 
In this way, the model can capture long-range dependencies within the data, as each token can attend to all others, regardless of their distance in the input sequence.
This approach allows ViTs to capture global dependencies between patches early in the network, making them effective for tasks requiring global context. 

Besides ViTs architectures based on \cite{dosovitskiy2020image}, in this paper we also explore the following variations.
\subsubsection{DeiT}
The Data-efficient Image Transformer (DeiT), introduced by \citep{touvron2021training}, is an improvement on ViT that focuses on reducing the data requirements for training Vision Transformers. While ViT performs well with large datasets, DeiTs introduce techniques such as knowledge distillation to train the model more efficiently on smaller datasets without sacrificing performances. DeiTs retain the transformer-based architecture of ViTs but optimize their training process, making it more accessible for practical applications where large amounts of labeled data may not be available.

%\subsubsection{Swin Transformer}
%The Swin Transformer, introduced by \citep{liu2021swin}, is a hierarchical vision transformer that enhances the scalability of ViTs by introducing local self-attention within non-overlapping windows. This window-based attention mechanism enables the model to capture both local and global features, similar to how CNNs operate on local receptive fields. The hierarchical structure of Swin Transformer allows it to handle images at different scales efficiently and makes it adaptable for tasks beyond image classification, such as object detection and segmentation. The Swin Transformer has been praised for its flexibility and has become a popular choice for a wide range of vision tasks.

\subsubsection{BEiT}
The Bidirectional Encoder representation from Image Transformers (BEiT), introduced by \citep{bao2021beit}, is a self-supervised vision transformer model that extends the principles of BERT-style pre-training from natural language processing to vision tasks. BEiT leverages masked image modeling, where portions of the input image patches are masked and the model is trained to predict the missing content based on the surrounding context. This pre-training technique allows BEiT to learn rich image representations without the need for extensive labeled data, making it highly effective in scenarios with limited supervision. After pre-training, BEiT can be fine-tuned for various downstream tasks, such as image classification, object detection, and segmentation. BEiT’s success lies in its ability to transfer knowledge learned during self-supervised pre-training to a wide range of vision applications, achieving competitive results while reducing the reliance on large labeled datasets.

\section{Results}

\begin{figure}
    \centering
    \includegraphics[width=0.5\textwidth]{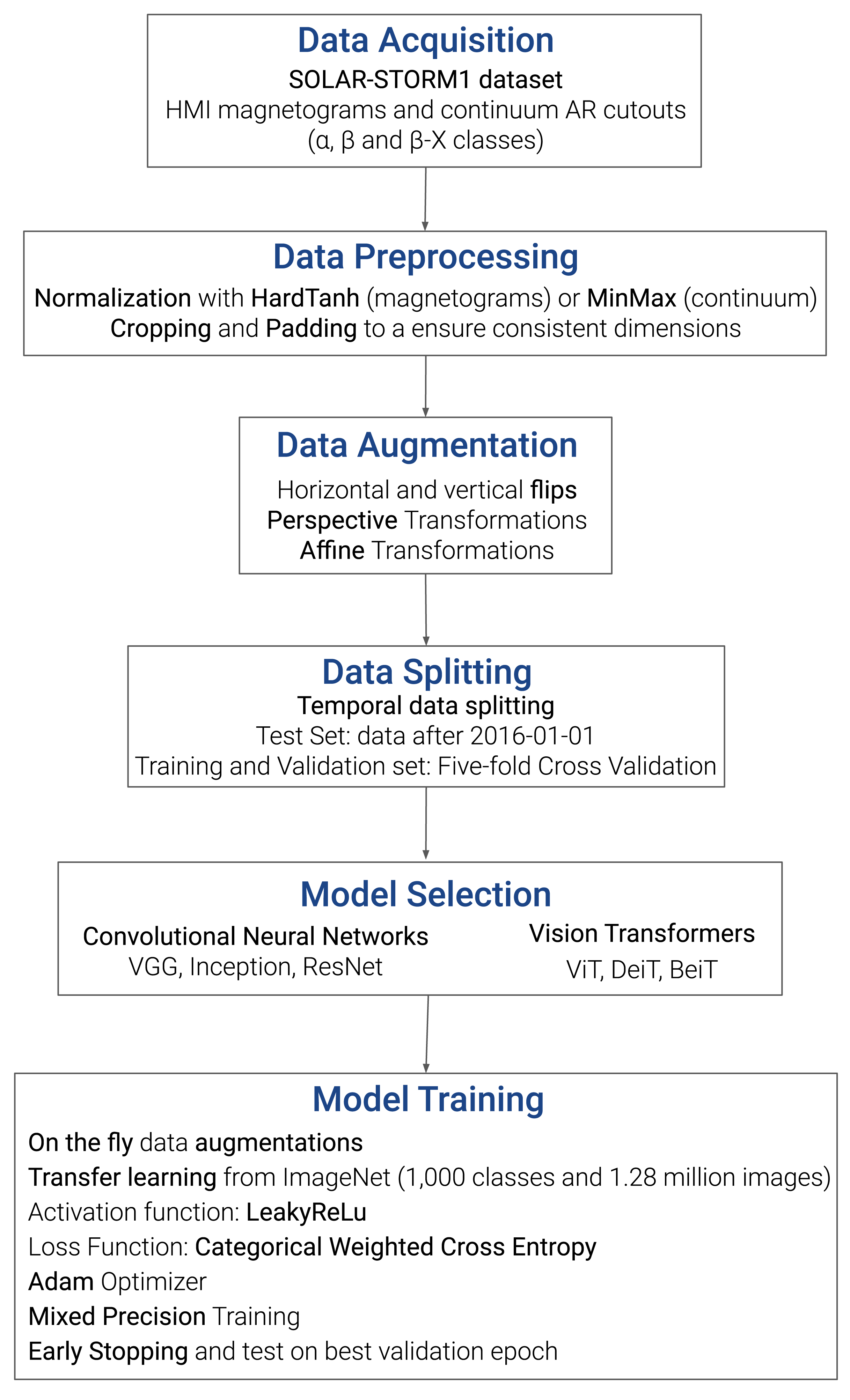}
    \caption{A summary of the workflow followed in this study.}
    \label{fig:workflow}
\end{figure}

The results presented in this section have been obtained by applying the pipeline summarized in Figure \ref{fig:workflow} to all the networks previously illustrated. To train these networks, we used the Adam optimizer \citep{Kingma14} with a learning rate of $10^{-5}$, employing a weighted categorical cross-entropy loss function to address class imbalance in the different folds. Once normalized, the images have been resized to a $224\times224$ square, with zero padding added to maintain the original aspect ratio. This was done so that the weights of the neural networks are initialized using transfer learning from pre-trained models on the ImageNet dataset \citep{deng2009imagenet}. 
This approach leverages the feature extraction capabilities learned from a large and diverse set of images, allowing the models to start with a strong foundation rather than from scratch. Even though transfer learning performs better when the source and target domains are closely related, it still offers significant advantages by improving convergence and potential accuracy. Additionally, early stopping was implemented to halt training if the validation loss did not improve. 
The model weights were then selected based on the epoch that achieved the best validation performance. To enhance the statistical reliability, the results are presented as the average test scores across all folds, along with their corresponding minimum and maximum values. Models have been trained on magnetogram data, continuum data and both combined in a 2 channel image using 2D convolutions. The number of trainable parameters for each model is reported in Table \ref{tab: trainable_parameters}. The results of the analysis are summarized in Table \ref{tab: results}.

\begin{table}[]
    \centering
    \begin{tabular}{cc}
    \toprule
    Model & Trainable Parameters (Millions) \\ 
    \midrule
    ResNet10t      & 4.92 \\ 
    ResNet18       & 11.2 \\ 
    ResNet26       & 14.9 \\ 
    Inception V4   & 41.1 \\ 
    VGG11          & 128.8 \\ 
    ViT Base       & 85.7 \\ 
    DEiT Base      & 85.7 \\ 
    BEiT Base      & 85.7 \\ 
    \bottomrule
    \end{tabular}
    \caption{Approximate value of trainable parameters for the models considered in this study. The number of trainable parameters is computed by iterating over all parameters in each model and summing the number of elements involved in the computation of gradients during backpropagation. The models trained from the timm library \citep{rw2019timm} are: resnet10t, resnet18, resnet26, inception\_v4, vgg11, vit\_base\_patch16\_224, deit\_base\_patch16\_224, beit\_base\_patch16\_224.}
    \label{tab: trainable_parameters}
\end{table}

\begin{table}[]
    \centering
    \tiny{
    \begin{tabular}{ccccccccc}
    \toprule
    Model & image type & Acc. (min-max-$\Delta$) & F1 Score (min-max-$\Delta$) & F1 $\alpha$ (min-max-$\Delta$) & F1 $\beta$ (min-max-$\Delta$) & F1 $\beta$-X (min-max-$\Delta$) \\ 
    \hline
 %   \midrule \multirow{3}{*}{ResNet10t} & magnetogram & 0.893 (0.871-0.904-0.033) & 0.849 (0.837-0.863-0.026) & 0.949 (0.918-0.963-0.045) & 0.872 (0.852-0.885-0.033) & 0.726 (0.697-0.748-0.051) \\
  ResNet10t & magnetogram & 0.893 (0.871-0.904-0.033) & 0.849 (0.837-0.863-0.026) & 0.949 (0.918-0.963-0.045) & 0.872 (0.852-0.885-0.033) & 0.726 (0.697-0.748-0.051) \\
     & continuum & 0.877 (0.845-0.893-0.049) & 0.821 (0.779-0.838-0.059) & 0.956 (0.946-0.961-0.015) & 0.848 (0.797-0.872-0.074) & 0.659 (0.586-0.686-0.100) \\
     & both & 0.894 (0.870-0.916-0.046) & 0.847 (0.816-0.877-0.061) & 0.953 (0.940-0.963-0.023) & 0.874 (0.849-0.902-0.054) & 0.715 (0.653-0.769-0.116) \\
    {ResNet18} & magnetogram & 0.887 (0.876-0.897-0.020) & 0.843 (0.825-0.858-0.033) & 0.949 (0.941-0.957-0.016) & 0.863 (0.845-0.876-0.031) & 0.718 (0.676-0.759-0.083) \\
     & continuum & 0.872 (0.853-0.887-0.033) & 0.825 (0.795-0.840-0.045) & 0.943 (0.925-0.949-0.025) & 0.842 (0.808-0.863-0.055) & 0.689 (0.628-0.709-0.081) \\
     & both & 0.903 (0.887-0.921-0.033) & 0.864 (0.845-0.891-0.046) & 0.953 (0.941-0.960-0.019) & 0.885 (0.868-0.906-0.038) & 0.756 (0.722-0.812-0.090) \\
    {ResNet26} & magnetogram & 0.882 (0.844-0.909-0.065) & 0.841 (0.800-0.883-0.083) & 0.940 (0.915-0.956-0.041) & 0.859 (0.817-0.895-0.078) & 0.724 (0.667-0.812-0.145) \\
     & continuum & 0.865 (0.817-0.904-0.087) & 0.823 (0.777-0.869-0.092) & 0.925 (0.888-0.949-0.061) & 0.840 (0.782-0.886-0.103) & 0.705 (0.662-0.774-0.112) \\
     & both & 0.883 (0.852-0.916-0.064) & 0.837 (0.788-0.881-0.092) & 0.945 (0.934-0.959-0.025) & 0.861 (0.823-0.900-0.077) & 0.704 (0.606-0.782-0.176) \\
    {Inception V4} & magnetogram & 0.897 (0.883-0.903-0.020) & 0.856 (0.843-0.864-0.021) & 0.948 (0.935-0.961-0.027) & 0.880 (0.866-0.887-0.021) & 0.739 (0.722-0.758-0.036) \\
     & continuum & 0.866 (0.838-0.892-0.054) & 0.817 (0.780-0.851-0.070) & 0.937 (0.924-0.948-0.023) & 0.838 (0.799-0.872-0.073) & 0.677 (0.608-0.735-0.127) \\
     & both & 0.854 (0.839-0.863-0.025) & 0.800 (0.773-0.819-0.046) & 0.919 (0.904-0.947-0.044) & 0.832 (0.812-0.849-0.036) & 0.649 (0.584-0.702-0.118) \\
    {VGG11} & magnetogram & 0.894 (0.868-0.925-0.057) & 0.843 (0.807-0.882-0.076) & 0.950 (0.922-0.966-0.044) & 0.875 (0.853-0.912-0.059) & 0.706 (0.644-0.769-0.126) \\
     & continuum & 0.875 (0.840-0.896-0.055) & 0.829 (0.798-0.850-0.053) & 0.938 (0.907-0.958-0.052) & 0.851 (0.816-0.874-0.059) & 0.696 (0.672-0.724-0.052) \\
     & both & 0.890 (0.858-0.910-0.051) & 0.832 (0.794-0.871-0.077) & 0.947 (0.933-0.959-0.026) & 0.873 (0.833-0.896-0.062) & 0.677 (0.583-0.765-0.182) \\
    {ViT Base} & magnetogram & 0.882 (0.871-0.896-0.025) & 0.835 (0.814-0.855-0.042) & 0.942 (0.916-0.956-0.040) & 0.859 (0.834-0.875-0.042) & 0.704 (0.651-0.750-0.099) \\
     & continuum & 0.848 (0.794-0.905-0.112) & 0.803 (0.740-0.866-0.126) & 0.919 (0.875-0.953-0.078) & 0.817 (0.744-0.885-0.141) & 0.673 (0.571-0.761-0.190) \\
     & both & 0.889 (0.863-0.915-0.052) & 0.845 (0.803-0.869-0.066) & 0.945 (0.918-0.959-0.040) & 0.867 (0.817-0.903-0.086) & 0.723 (0.633-0.765-0.133) \\
    {DEiT Base} & magnetogram & 0.901 (0.885-0.916-0.031) & 0.861 (0.840-0.885-0.045) & 0.952 (0.933-0.968-0.035) & 0.881 (0.856-0.901-0.045) & 0.749 (0.700-0.802-0.102) \\
     & continuum & 0.864 (0.834-0.892-0.059) & 0.817 (0.778-0.848-0.070) & 0.932 (0.911-0.951-0.041) & 0.838 (0.789-0.872-0.083) & 0.680 (0.611-0.723-0.112) \\
     & both & 0.902 (0.876-0.926-0.049) & 0.853 (0.812-0.888-0.076) & 0.957 (0.925-0.968-0.043) & 0.884 (0.865-0.912-0.048) & 0.718 (0.645-0.786-0.142) \\
    {BEiT Base} & magnetogram & 0.873 (0.854-0.895-0.041) & 0.828 (0.801-0.856-0.055) & 0.937 (0.917-0.950-0.033) & 0.846 (0.819-0.874-0.055) & 0.702 (0.646-0.748-0.102) \\
     & continuum & 0.866 (0.842-0.902-0.060) & 0.817 (0.786-0.860-0.074) & 0.938 (0.918-0.952-0.034) & 0.838 (0.801-0.883-0.081) & 0.676 (0.616-0.745-0.129) \\
     & both & 0.891 (0.844-0.920-0.076) & 0.845 (0.772-0.886-0.114) & 0.942 (0.907-0.961-0.054) & 0.873 (0.831-0.904-0.073) & 0.720 (0.577-0.793-0.216) \\
    \bottomrule
    \end{tabular}
    }
    \caption{Performance results for different deep learning models. 
    The table reports the mean values of the metrics across the five folds, along with minimum and maximum values and the difference $\Delta$ between maximum and minimum values. 
    Model names are the same as in Table \ref{tab: trainable_parameters}.}
    \label{tab: results}
\end{table}

Among all the models, ResNet18 trained on both magnetogram and continuum images achieved the highest average accuracy ($0.903$) and highest average F1 score ($0.864$), demonstrating the efficacy of combining both image types in providing richer information for model training. This suggests that the ResNet architecture can effectively leverage diverse inputs to outperform other architectures in terms of overall classification accuracy.

When analyzing models trained on only magnetograms, the DEiT Base model stands out, achieving the highest average accuracy ($0.901$) and highest F1 score of 0.861. This indicates that transformer-based architectures can capture relevant spatial features present in magnetograms (even if ResNet-based architectures remain competitive also in this case). For models trained exclusively on continuum images, ResNet10t achieved the highest average accuracy ($0.877$) and ResNet18 the highest average F1 score ($0.825$). This suggests that continuum images may contain less spatial variability than magnetograms, and a simple ResNet-based architecture with relatively few parameters is sufficient to extract meaningful patterns from the data. In general, accuracy on continuum images tends to be slightly lower across all models compared to magnetograms or combined image types, likely due to the more homogeneous nature of this data type.

When analyzing the F1 score for the $\alpha$ class, ResNet10t and ResNet18 achieve the highest average scores ($0.953$) when trained on both image types, while for the $\beta$ class the highest F1 score on average is obtained with the ResNet18 network. Models trained only on magnetograms, such as ResNet18 and DEiT Base, also performed well, achieving F1 $\beta$ scores of 0.876 and 0.881, respectively. However, models trained solely on continuum data, like ViT Base (F1 $\beta$ = 0.817), showed a noticeable drop in performance, possibly due to the limited variability in continuum images. In terms of the F1 $\beta$-X score, which evaluates performance in the most challenging class, as expected the DEiT architecture achieves the best average score ($0.749$), suggesting that transformer-based models are better suited for capturing the intricate spatial relationships present in complex magnetic structures. However, performance declines notably when models are trained exclusively on continuum images.

The observed variability across folds, indicated by $\Delta$ values, generally ranges from 2\% to 10\% for accuracy and overall F1 scores. 
For the more complex $\beta$-X class, the variation can be even higher, with differences of up to 20\%. Anyway, in general, the average performances do not change drastically from one model architecture to another. This could be attributed to several factors. First, complex models may not fully utilize their capacity if the dataset does not present sufficient complexity or variability, leading to comparable performances with simpler architectures that can efficiently capture the essential patterns. Additionally, simpler models benefit from reduced risk of overfitting due to their smaller number of parameters, particularly when trained on datasets with moderate complexity. This is further enhanced by the use of modern training techniques such as batch normalization, dropout, and advanced optimizers, which improve generalization across models. These factors collectively explain why more complex architectures may not consistently outperform simpler ones, particularly if a specific dataset does not demand the added complexity. This closeness can also in part be attributed to the extensive data augmentation procedure employed during the training phase.
In particular, ViTs tend to be less sensitive to data augmentation compared to traditional CNNs, mainly due to their global attention mechanism and ability to model relationships across the entire image. Transformers inherently capture more comprehensive image representations, making them more robust to small variations that data augmentations typically address in CNNs. As a result, extensive data augmentations help to narrow the performance gap between CNNs and more complex transformer models.

Compared to similar works, performance metrics reported by the study of \cite{fang2019deep} are not directly comparable due to the random splitting of a highly redundant dataset.
Comparison with the study of \cite{tang2021multiple}, where 3D convolutions are used to classify an image consisting of both magnetogram and continuum data, shows that the maximum scores obtained with 2D convolutions are comparable to a more involved ensemble model. However, our goal was not to focus on the individual absolute values of the scores in comparison to other works, as it is possible to continue training with different random seeds in search of a favourable outlier that yields higher scores \citep{picard2021torch}. 
Instead, we aimed to provide a meaningful comparison of the performance of different model architectures under similar conditions, highlighting their relative strengths and weaknesses from a global perspective.

\section{Conclusions}
In this study, we analyzed different deep learning architectures for the task of classifying the magnetic class of solar cutouts.
We observed that combining magnetogram and continuum image types enhances model performance by leveraging complementary features from diverse inputs.
When considering only magnetograms, data efficient transformer models achieve the best performance, demonstrating the effectiveness of transformer-based architectures in capturing the spatial complexity of magnetograms.
Models trained exclusively on continuum images exhibit overall lower performance, suggesting that continuum images, due to their more homogeneous nature, offer less spatial variability.

Overall, performance metrics do not change drastically between model architectures, a fact that can be attributed to several factors: complex models may not fully utilize their capacity if the dataset lacks sufficient complexity or variability, resulting in performance comparable to simpler architectures that can effectively capture the essential patterns; 
simpler models have a reduced risk of overfitting, particularly when trained on moderately complex datasets; modern training techniques such as batch normalization, dropout, and advanced optimizers further improve generalization, helping to reduce the performance gap between simpler and more complex models. Finally, the extensive data augmentation procedures employed during the training phase also contribute to closing the gap between simpler CNNs and more complex ViTs.

\section*{Acknowledgement}
\label{sec:acknowledge}
EL and AMM acknoledge the HORIZON Europe ARCAFF Project, Grant No. 101082164. 
SG, MP and AMM acknowledge INdAM-GNCS;
SG was supported by the Programma Operativo Nazionale (PON) “Ricerca e Innovazione” 2014–2020. 

\bibliography{refs}{}

\begin{thebibliography}{}
\expandafter\ifx\csname natexlab\endcsname\relax\def\natexlab#1{#1}\fi
\providecommand{\url}[1]{\href{#1}{#1}}
\providecommand{\dodoi}[1]{doi:~\href{http://doi.org/#1}{\nolinkurl{#1}}}
\providecommand{\doeprint}[1]{\href{http://ascl.net/#1}{\nolinkurl{http://ascl.net/#1}}}
\providecommand{\doarXiv}[1]{\href{https://arxiv.org/abs/#1}{\nolinkurl{https://arxiv.org/abs/#1}}}

\bibitem[{Abd {et~al.}(2010)Abd, Majed, \& Zharkova}]{abd2010automated}
Abd, M.~A., Majed, S.~F., \& Zharkova, V. 2010, in Technological developments
  in networking, education and automation, Springer, 321--325

\bibitem[{Baker(2002)}]{baker2002cope}
Baker, D.~N. 2002, Science, 297, 1486

\bibitem[{Bao {et~al.}(2021)Bao, Dong, Piao, \& Wei}]{bao2021beit}
Bao, H., Dong, L., Piao, S., \& Wei, F. 2021, arXiv preprint arXiv:2106.08254

\bibitem[{Benz(2017)}]{benz2017flare}
Benz, A.~O. 2017, Living reviews in solar physics, 14, 1

\bibitem[{Buzulukova \& Tsurutani(2022)}]{buzulukova2022space}
Buzulukova, N., \& Tsurutani, B. 2022, Frontiers in Astronomy and Space
  Sciences, 9, 1017103

\bibitem[{Cerqueira {et~al.}(2024)Cerqueira, Santos, Baghoussi, \&
  Soares}]{cerqueira2024fly}
Cerqueira, V., Santos, M., Baghoussi, Y., \& Soares, C. 2024, arXiv preprint
  arXiv:2404.16918

\bibitem[{Colak \& Qahwaji(2008)}]{colak2008automated}
Colak, T., \& Qahwaji, R. 2008, Solar Physics, 248, 277

\bibitem[{Deng {et~al.}(2009)Deng, Dong, Socher, Li, Li, \&
  Fei-Fei}]{deng2009imagenet}
Deng, J., Dong, W., Socher, R., {et~al.} 2009, in 2009 IEEE conference on
  computer vision and pattern recognition, Ieee, 248--255

\bibitem[{Dosovitskiy(2020)}]{dosovitskiy2020image}
Dosovitskiy, A. 2020, arXiv preprint arXiv:2010.11929

\bibitem[{Eren {et~al.}(2016)Eren, Kilcik, Atay, Miteva, Yurchyshyn, Rozelot,
  \& Ozguc}]{eren2016flare}
Eren, S., Kilcik, A., Atay, T., {et~al.} 2016, Monthly Notices of the Royal
  Astronomical Society, stw2742

\bibitem[{Fang {et~al.}(2019)Fang, Cui, Ao, {et~al.}}]{fang2019deep}
Fang, Y., Cui, Y., Ao, X., {et~al.} 2019, Advances in Astronomy

\bibitem[{Georgoulis {et~al.}(2024)Georgoulis, Yardley, Guerra, Murray,
  Ahmadzadeh, Anastasiadis, Angryk, Aydin, Banerjee, Barnes,
  {et~al.}}]{georgoulis2024prediction}
Georgoulis, M.~K., Yardley, S.~L., Guerra, J.~A., {et~al.} 2024, Advances in
  Space Research

\bibitem[{Goodfellow(2016)}]{goodfellow2016deep}
Goodfellow, I. 2016, Deep learning, Vol. 196 (MIT press)

\bibitem[{Guastavino {et~al.}(2022)Guastavino, Marchetti, Benvenuto, Campi, \&
  Piana}]{guastavino2022implementation}
Guastavino, S., Marchetti, F., Benvenuto, F., Campi, C., \& Piana, M. 2022,
  Astronomy \& Astrophysics, 662, A105

\bibitem[{Guastavino {et~al.}(2023)Guastavino, Marchetti, Benvenuto, Campi, \&
  Piana}]{guastavino2023operational}
---. 2023, Frontiers in Astronomy and Space Sciences, 9,

\bibitem[{He {et~al.}(2016)He, Zhang, Ren, \& Sun}]{he2016deep}
He, K., Zhang, X., Ren, S., \& Sun, J. 2016, in Proceedings of the IEEE
  conference on computer vision and pattern recognition, 770--778

\bibitem[{Hudson(2011)}]{hudson2011global}
Hudson, H.~S. 2011, Space Science Reviews, 158, 5

\bibitem[{Kingma \& Ba(2015)}]{Kingma14}
Kingma, D.~P., \& Ba, J. 2015, in 3rd International Conference on Learning
  Representations, {ICLR} 2015, San Diego, CA, USA, May 7-9, 2015, Conference
  Track Proceedings, ed. Y.~Bengio \& Y.~LeCun.

\bibitem[{Krizhevsky {et~al.}(2012)Krizhevsky, Sutskever, \&
  Hinton}]{krizhevsky2012imagenet}
Krizhevsky, A., Sutskever, I., \& Hinton, G.~E. 2012, Advances in neural
  information processing systems, 25

\bibitem[{LeCun {et~al.}(2015)LeCun, Bengio, \& Hinton}]{lecun2015deep}
LeCun, Y., Bengio, Y., \& Hinton, G. 2015, nature, 521, 436

\bibitem[{LeCun {et~al.}(1998)LeCun, Bottou, Bengio, \&
  Haffner}]{lecun1998gradient}
LeCun, Y., Bottou, L., Bengio, Y., \& Haffner, P. 1998, Proceedings of the
  IEEE, 86, 2278

\bibitem[{McCloskey {et~al.}(2016)McCloskey, Gallagher, \&
  Bloomfield}]{mccloskey2016flaring}
McCloskey, A.~E., Gallagher, P.~T., \& Bloomfield, D.~S. 2016, Solar Physics,
  291, 1711

\bibitem[{McCloskey {et~al.}(2018)McCloskey, Gallagher, \&
  Bloomfield}]{mccloskey2018flare}
---. 2018, Journal of Space Weather and Space Climate, 8, A34

\bibitem[{Pesnell {et~al.}(2012)Pesnell, Thompson, \&
  Chamberlin}]{pesnell2012solar}
Pesnell, W., Thompson, B., \& Chamberlin, P. 2012, The solar dynamics
  observatory (SDO) (Springer)

\bibitem[{Piana {et~al.}(2022)Piana, Emslie, Massone, \&
  Dennis}]{piana2022hard}
Piana, M., Emslie, A.~G., Massone, A.~M., \& Dennis, B.~R. 2022, Hard X-Ray
  Imaging of Solar Flares, Vol. 164 (Springer)

\bibitem[{Picard(2021)}]{picard2021torch}
Picard, D. 2021, arXiv preprint arXiv:2109.08203

\bibitem[{Scherrer {et~al.}(2012)Scherrer, Schou, Bush, Kosovichev, Bogart,
  Hoeksema, Liu, Duvall, Zhao, Title, {et~al.}}]{scherrer2012helioseismic}
Scherrer, P.~H., Schou, J., Bush, R., {et~al.} 2012, Solar Physics, 275, 207

\bibitem[{Simonyan \& Zisserman(2014)}]{simonyan2014very}
Simonyan, K., \& Zisserman, A. 2014, arXiv preprint arXiv:1409.1556

\bibitem[{Szegedy {et~al.}(2015)Szegedy, Liu, Jia, Sermanet, Reed, Anguelov,
  Erhan, Vanhoucke, \& Rabinovich}]{szegedy2015going}
Szegedy, C., Liu, W., Jia, Y., {et~al.} 2015, in Proceedings of the IEEE
  conference on computer vision and pattern recognition, 1--9

\bibitem[{Tang {et~al.}(2021)Tang, Zeng, Chen, Liao, Wang, Luo, Chen, Cui,
  Zhou, Deng, {et~al.}}]{tang2021multiple}
Tang, R., Zeng, X., Chen, Z., {et~al.} 2021, The Astrophysical Journal
  Supplement Series, 257, 38

\bibitem[{Touvron {et~al.}(2021)Touvron, Cord, Douze, Massa, Sablayrolles, \&
  J{\'e}gou}]{touvron2021training}
Touvron, H., Cord, M., Douze, M., {et~al.} 2021, in International conference on
  machine learning, PMLR, 10347--10357

\bibitem[{Vaswani(2017)}]{vaswani2017attention}
Vaswani, A. 2017, Advances in Neural Information Processing Systems

\bibitem[{Wightman(2019)}]{rw2019timm}
Wightman, R. 2019, PyTorch Image Models,
 
\end{thebibliography}
\bibliographystyle{aasjournal}

\end{document}